\documentclass[runningheads]{svmult}

\usepackage{epsf}      
\usepackage{makeidx}   
\usepackage{graphicx}  
\usepackage{subeqnar}  
\usepackage{multicol}  
\usepackage{physprbb}  
\makeindex             

\newcommand{\sig}{\:\lower0.6ex\hbox{$\stackrel{\textstyle >}{\sim}$}\:}
\newcommand{\sil}{\:\lower0.6ex\hbox{$\stackrel{\textstyle <}{\sim}$}\:}
\newcommand{\sigs}{\:\lower0.4ex\hbox{$\stackrel{\scriptstyle
      >}{\scriptstyle \sim}$}\,}
\newcommand{\sils}{\:\lower0.4ex\hbox{$\stackrel{\scriptstyle
      <}{\scriptstyle \sim}$}\,}
\begin{document}
\title*{Mass Spectra from Turbulent Fragmentation}
%
%
%
%
\titlerunning{Mass Spectra from Turbulent Fragmentation}
%
\author{Ralf S.\ Klessen\inst{1,2}}
\authorrunning{Ralf S.\ Klessen}
%
%
\institute{UCO/Lick Observatory, University of California at
 Santa Cruz, Santa
 Cruz, CA 95064, USA
\and Max-Planck-Institut f{\"u}r
 Astronomie, K{\"o}nigstuhl 17, 69117 Heidelberg, Germany}

\maketitle              

\begin{abstract}
      Turbulent fragmentation determines where and when protostellar
      cores form, and how they contract and grow in mass from the
      surrounding cloud material. This process is investigated, using
      numerical models of molecular cloud dynamics. Molecular cloud
      regions without turbulent driving sources, or where turbulence
      is driven on large scales, exhibit rapid and efficient star
      formation in a clustered mode, whereas interstellar turbulence
      that carries most energy on small scales results in isolated
      star formation with low efficiency. The clump mass spectrum of
      shock-generated density fluctuations in pure hydrodynamic,
      supersonic turbulence is not well fit by a power law, and it is
      too steep at the high-mass end to be in agreement with the
      observational data. When gravity is included in the turbulence
      models, local collapse occurs, and the spectrum extends towards
      larger masses as clumps merge together, a power-law description
      $dN/dM \propto M^{\nu}$ becomes possible with slope $\nu \sils
      -2$. In the case of pure gravitational contraction, i.e. in
      regions without turbulent support, the clump mass spectrum is
      shallower with $\nu \approx -3/2$. The mass spectrum of
      protostellar cores in regions without turbulent support and
      where turbulence is replenished on large-scales, however, is
      well described by a log-normal or by multiple power laws,
      similar to the stellar IMF at low and intermediate masses. In
      the case of small-scale turbulence, the core mass spectrum is
      too flat compared to the IMF for all masses.
\end{abstract}

\section{Introduction}
\label{sec:intro}
Stars are born in turbulent interstellar clouds of molecular hydrogen.
The location and the mass growth of young stars are hereby intimately
coupled to the dynamical cloud environment. Stars form by
gravitational collapse of shock-compressed density fluctuations
generated from the supersonic turbulence ubiquitously observed in
molecular clouds (e.g.\ Elmegreen 1993, Padoan 1995, Klessen, Heitsch,
\& Mac~Low 2000). Once a gas clump becomes gravitationally unstable,
it begins to collapse and the central density increases considerably,
giving birth to a protostar.  As stars typically form in small
aggregates or larger clusters (Lada 1992, Adams \& Myers 2001) the
interaction of protostellar cores and their competition for mass
growth from their surroundings are important processes determining the
distribution of stellar masses. These complex phenomena are addressed
by analyzing and comparing numerical models of self-gravitating
supersonic turbulence. Focus is on the connection between supersonic
turbulence and local collapse and on the relation between the mass
spectra of transient gas clumps and protostellar cores.

\section{Numerical Models of Turbulent Molecular Clouds}
\label{sec:numerics}
To adequately describe turbulent fragmentation and the formation of
protostellar cores, it is necessary to resolve the collapse of shock
compressed regions over several orders of magnitude in
density. To achieve this, I use SPH ({\em smoothed particle
hydrodynamics}) which is a well-known particle-based Lagrangian method
to solve the equations of hydrodynamics. Details about the numerical
implementation can be found in Klessen \& Burkert (2000).

In the current paper I consider different scenarios for interstellar
turbulence. Model 1 completely lacks turbulent support (it describes
the contraction of Gaussian density distribution, Klessen \& Burkert
2000, 2001), model 2 describes freely decaying supersonic turbulence,
and in models 3 to 5 turbulence is continuously driven. The non-local
driving scheme inserts kinetic energy in a specified range of
wavenumbers, $1 \le k \le 2$, $3 \le k \le 4$, and $7 \le k \le 8$,
corresponding to sources that act on large, intermediate, and small
scales, respectively (Mac Low 1999, Klessen et al.\ 2000), such that an
rms Mach numer of ${\cal M}_{\rm rms} = 5.5$ is maintained throughout
the evolution.

\section{Star Formation from Turbulent Fragmentation}
\label{sec:location-time}
Stars form from turbulent fragmentation of molecular cloud material.
Supersonic turbulence, even if strong enough to counterbalance
gravity on global scales, will usually {\em provoke} local collapse.
Turbulence establishes a complex network of interacting shocks, where
converging shock fronts generate clumps of high density. This density
enhancement can be large enough for the fluctuations to become
gravitationally unstable and collapse.  However, the fluctuations in
turbulent velocity fields are highly transient and the same random
flow that creates local density enhancements can disperse them again.
For local collapse to actually result in the formation of stars,
Jeans-unstable shock-generated density fluctuations must collapse to
sufficiently high densities on time scales shorter than the typical
time interval between two successive shock passages.  The shorter the
time between shock passages, the less likely these fluctuations are to
survive. Hence, the timescale and efficiency of protostellar core
formation depend strongly on the wavelength and strength of the
driving source (Klessen et al.\ 2000, Heitsch et al.\ 2001), and
accretion histories of individual protostars are strongly time varying
(Klessen 2001).
\begin{figure*}[pt]
\begin{center}
\hspace{-1.7cm}
\includegraphics[width=0.95\textwidth]{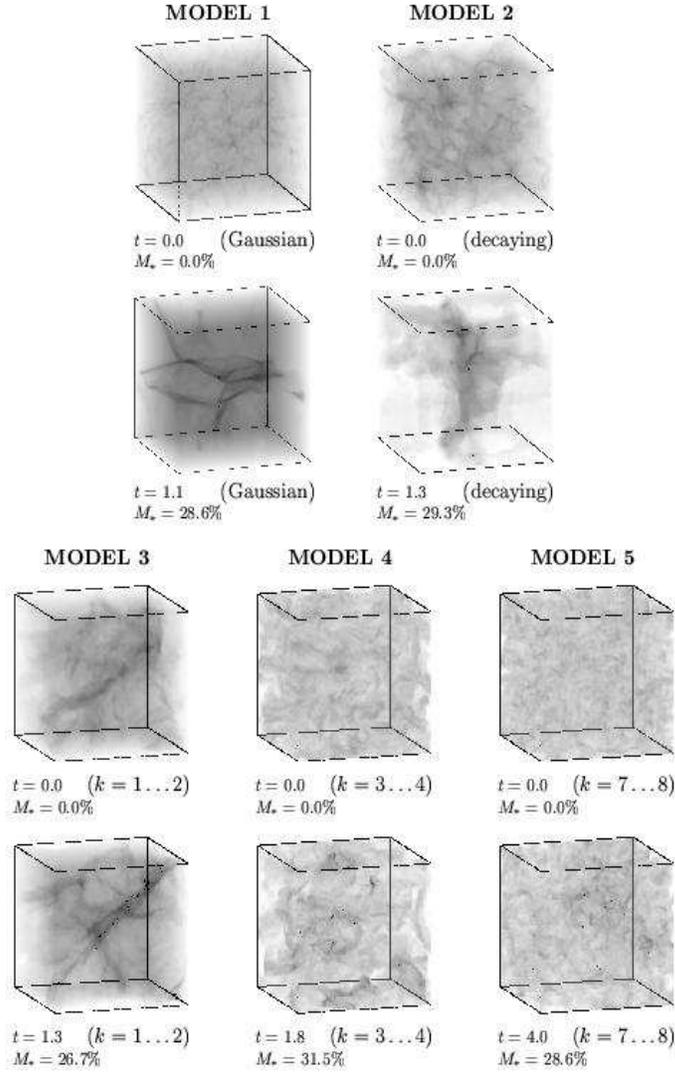}
\vspace{-0.7cm}
\end{center}
\caption{\label{fig:3Dplot} Comparison of the
  gas distribution in the five models at two different phases of the
  dynamical evolution, at $t=0$ indicating the initial density
  structure, just before gravity is `switched on', and after the first
  cores have formed and accumulated roughly $M_{\rm *} \approx 30$\%
  of the total mass. The high-density (protostellar) cores are
  indicated by black dots. Note the different time interval needed to
  reach the same dynamical stage. Time is normalized to the global
  free-fall timescale of the system, which is $\tau_{\rm ff} =
  10^5\,$yr for $T=11.4\,$K and $n({\rm H}_2) = 10^5\,$cm$^{-3}$. The
  cubes contain masses of $220\,\langle M_{\rm J}\rangle$ (models 1
  and 2) and $120\,\langle M_{\rm J}\rangle$
  (models 3 to 5), respectively, where the average thermal Jeans mass
  is $\langle M_{\rm J}\rangle= 1\,$M$_{\odot}$ with the above scaling. The
  considered volumes are $(0.32\,$pc$)^3$ and $(0.29\,$pc$)^3$,
  respectively. }
\end{figure*}

The velocity field of long-wavelength turbulence is dominated by
large-scale shocks which are very efficient in sweeping up molecular
cloud material, thus creating massive coherent density
structures. When a coherent region reaches the critical density for
gravitational collapse its mass typically exceeds the local Jeans
limit by far.  Inside the shock compressed region, the velocity
dispersion is much smaller than in the ambient turbulent flow and the
situation is similar to localized tur\-bulent decay.  Quickly a
cluster of protostellar cores forms. Therefore, models 1 to 3 lead to
a {\em clustered} mode of star formation. The efficiency of turbulent
fragmentation is reduced if the driving wavelength decreases. When
energy is inserted mainly on small spatial scales, the network of
interacting shocks is very tightly knit, and protostellar cores form
independently of each other at random locations throughout the cloud
and at random times.  Individual shock generated clumps have lower
mass and the time interval between two shock passages through the same
point in space is small.  Hence, collapsing cores are easily destroyed
again and star formation is inefficient. This scenario corresponds to
the {\em isolated} mode of star formation.

This is visualized in Figure~\ref{fig:3Dplot}, showing the density
structure of all five models initially and at a time when the first
protostellar cores have formed by turbulent fragmentation and have
accreted roughly 30\% of the total mass.  Note the different time
interval needed to reach this dynamical state and the large
variations in the resulting density distribution between the various
models.  Dark dots indicate the location of dense collapsed cores.

\section{Mass Spectra from Turbulent Fragmentation}
\label{sec:mass-spectra}
\begin{figure*}[tp]
\begin{center}
\vspace{-2.3cm}
\includegraphics[width=.95\textwidth]{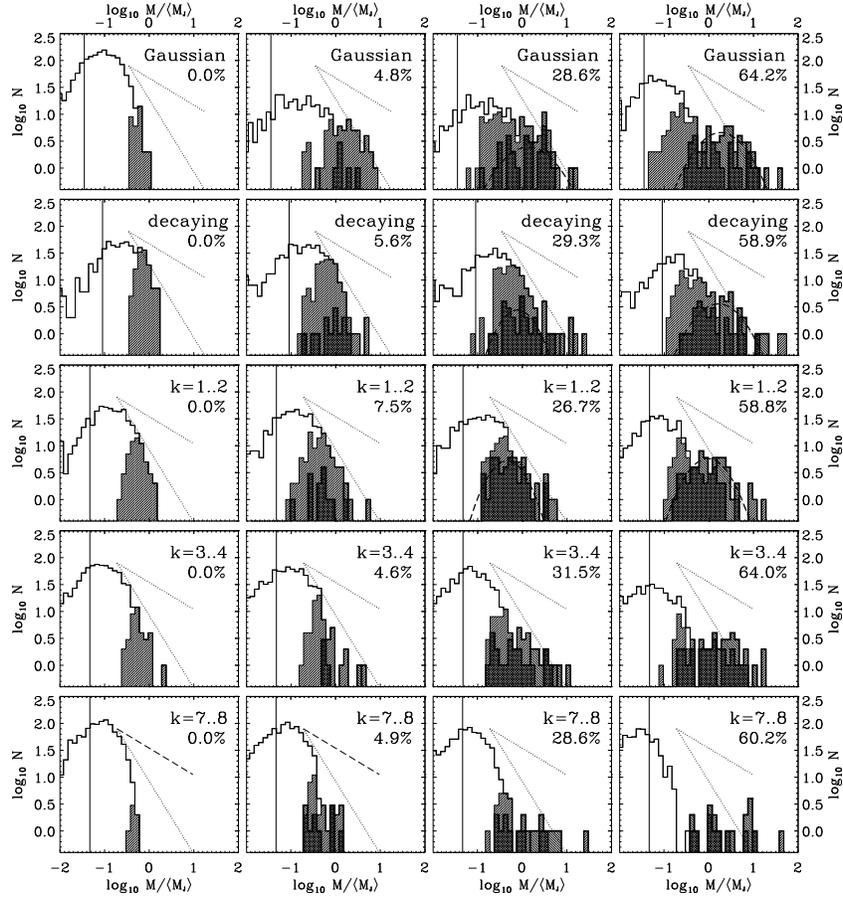}
\vspace{-1.3cm}
\end{center}
\caption{\label{fig:massspectra} Mass
    spectra of gas clumps (thin lines), and of the subset of Jeans
    unstable clumps (thin lines, hatched distribution), and of dense
    collapsed cores (hatched thick-lined histograms).  Masses are
    binned logarithmically and normalized to the average thermal Jeans
    mass $\langle M_{\rm J}\rangle$. The left column gives the initial
    state of the system, just when gravity is `switched on', the
    second column shows the mass spectra when $M_{\rm *} \approx 5$\%
    of the mass is accreted onto dense cores, the third column
    describes $M_{\rm *} \approx 30$\%, and the last one $M_{\rm *}
    \approx 60$\%. For comparison with power-law spectra ($dN/dM
    \propto M^{\nu}$), a slope $\nu = -1.5$ typical for the observed
    clump mass distribution, and the Salpeter slope $\nu=-2.33$ for
    the IMF at intermediate and large masses, are indicated by the
    dotted lines in each plot. Note that with the adopted logarithmic
    mass binning these slopes appear shallower by $+1$ in the plot.
    The vertical line shows the SPH resolution limit. In columns 3 and
    4, the long dashed curve shows the best log-normal fit to the core
    mass spectrum. To compare the distribution of core masses with the
    stellar IMF, an efficiency factor of roughly $1/3$ to $1/2$ for
    the conversion of protostellar core material into single stars
    needs to be taken into account. For
    $T=11.4\,$K and $n({\rm H}_2) = 10^5\,$cm$^{-3}$, the average
    Jeans mass in the system is $\langle M_{\rm J}\rangle=
    1\,$M$_{\odot}$. }
\end{figure*}
To illustrate the relation between the masses of molecular clumps,
protostellar cores and the resulting stars,  Figure 2 plots for the
five models the mass distribution of all gas clumps, of the subset of
Jeans-critical clumps, and of collapsed cores at four different
evolutionary stages -- clumps are defined as described in Appendix 1
of Klessen \& Burkert (2000).

In the initial, completely pre-stellar phase the clump mass spectrum
is not
well described by a single power law. The distribution has small width
and falls off steeply at large masses. Below masses $M \approx 0.3
\,\langle M_{\rm J} \rangle$ the distribution becomes shallower, and
strongly declines at and beyond the resolution limit (indicated by a
vertical line). Clumps are on average considerably smaller than the
mean Jeans mass in the system $\langle M_{\rm J}\rangle$.

Gravity strongly modifies the distribution of clump masses during the
later evolution. As gas clumps merge and grow bigger, their mass spectrum
becomes flatter and extends towards larger masses. Consequently, the
number of clumps that exceed the Jeans limit grows, and local collapse
sets in leading to the formation of dense condensations. This is most
evident in models 1 and 2 where the velocity field
is entirely determined by gravitational contraction on all scales. The
clump mass spectrum in intermediate phases of the evolution (i.e.\ 
when protostellar cores are forming, but the overall gravitational
potential is still dominated by non-accreted gas) exhibits a slope
$-1.5$ similar to the observed one. When the velocity field is
dominated by strong (driven) turbulence, the effect of gravity on the
clump mass spectrum is much weaker. It remains steep, close to or even
below the Salpeter value. This is most clearly seen for
small-wavelength turbulence. Here, the short interval between shock
passages prohibits efficient merging and the build up of a large
number of massive clumps. Only few fluctuations become Jeans unstable
and collapse to form protostars. These form independent of each other
at random locations and times and typically do not interact.
Increasing the driving wavelength leads to more coherent and rapid
star formation. 

Long-wavelength turbulence or turbulent decay leads to a core mass
spectrum that is well approximated by a {\em log-normal}. It roughly
peaks at the {\em average thermal Jeans mass} $\langle M_{\rm
  J}\rangle$ of the system and is comparable in width with the
observed IMF (Klessen \& Burkert 2000, 2001). The log-normal shape of
the mass distribution may be explained by invoking the central limit
theorem (e.g.\ Zinnecker 1984), as protostellar cores form and evolve
through a sequence of highly stochastic events (resulting from
supersonic turbulence and/or competitive accretion). To find the mass
peak at $\langle M_{\rm J}\rangle$ may be somewhat surprising given
the fact that the local Jeans mass strongly varies between different
clumps. In a statistical sense the system retains knowledge of its
mean properties.  The total width of the core distribution is about
two orders of magnitude in mass and is approximately the same for all
four models.  However, the spectrum for intermediate and
short-wavelength turbulence, i.e.\ for isolated core formation, is too
flat (or equivalently too wide) to be comparable to the observed IMF.
This is in agreement with the hypothesis that most stars form in
aggregates or clusters.

The current findings raise considerable doubts about attempts to
explain the stellar IMF from the turbulence-induced clump mass
spectrum {\em only} (e.g.\ Elmegreen 1993, Padoan 1995, Padoan \&
Nordlund 2001). Quite typically for star forming turbulence, the
collapse timescale of shock-compressed gas clumps often is comparable
to their lifetime (molecular cloud clumps appear to be very transient,
e.g.\ Bergin et al.\ 1997).  While collapsing to form or feed
protostars, clumps may loose or gain matter from interaction with the
ambient turbulent flow. In a dense cluster environment, collapsing
clumps may merge to form larger clumps containing multiple
protostellar cores, which subsequently compete with each other for
accretion form the common gas environment. As consequence, the
resulting distribution of clump masses in star forming regions
strongly evolves in time (Figure 2). It is not
possible to infer a {\em one-to-one} relation between the clump masses
resulting from turbulent molecular cloud fragmentation and the stellar
IMF. It is not appropriate to take a snapshot of the turbulent clump
mass spectrum as  describing the IMF.

%

%


\begin{thebibliography}{}
\addcontentsline{toc}{section}{References}

\bibitem{} Adams, F.\ C., Myers, P.\ C. 2001, ApJ, in press (astro-ph/0102039)
\bibitem{} Bergin, E.\ A., Goldsmith, P.\ F., Snell, R.\ L.,
  Langer, W.\ D. 1997, ApJ, 428, 285
\bibitem{} Elmegreen, B.\ G. 1993, ApJ, 419, L29
\bibitem{} Heitsch, F., Mac~Low, M.-M., Klessen, R.\ S. 2001, ApJ,
  547, 280
\bibitem{} Klessen, R.\ S. 2001, ApJ, 550, L77
\bibitem{} Klessen, R.\ S., Burkert, A. 2000, ApJS, 128, 287
\bibitem{} Klessen, R.\ S., Burkert, A. 2001, ApJ, 549, 386
\bibitem{} Klessen, R.\ S., Heitsch, F., Mac~Low, M.-M. 2000, ApJ,
  535, 887
\bibitem{} Lada, E., 1992, ApJ, 393, L25
\bibitem{} Mac~Low, M.-M., 1999, ApJ, 524, 169
\bibitem{} Mac~Low, M.-M., Klessen, R.\ S., Burkert, A., Smith, M.\ D.
1998, PRL, 80, 2754
\bibitem{} Padoan, P. 1995, MNRAS, 277, 337
\bibitem{} Padoan, P., Nordlund, \AA. 2001, ApJ, submitted (astro-ph/0011465)
\bibitem{} Zinnecker, H. 1984, MNRAS, 210, 43
\end{thebibliography}
\end{document}